\documentstyle[psfig,prb,aps]{revtex}
\begin{document}
\draft

\twocolumn[\hsize\textwidth\columnwidth\hsize\csname
@twocolumnfalse\endcsname

\title{
An orbital-free molecular dynamics study of melting in K$_{20}$, K$_{55}$,
K$_{92}$, K$_{142}$, Rb$_{55}$ and Cs$_{55}$ clusters.
}
\author{Andr\'es Aguado}
\address{Departamento de F\'\i sica Te\'orica,
Universidad de Valladolid, Valladolid 47011, Spain}
\maketitle
\begin{abstract}
The melting-like transition in potasium clusters K$_N$, with N=20, 55, 92 and 
142, is studied by using an orbital-free density-functional
constant-energy molecular dynamics simulation method, and compared to
previous theoretical results on the
melting-like transition in sodium clusters of the same sizes. Melting in
potasium and sodium clusters proceeds in a similar way:
a surface melting stage develops upon heating before the homogeneous melting
temperature is reached. Premelting effects are nevertheless more important and
more easily established in potasium clusters, and the transition regions spread
over temperature intervals which are wider than in the case of sodium. For all
the sizes considered, the percentage melting temperature reduction when passing
from Na to K clusters is substantially larger than in the bulk. Once
those two materials have been compared for a number of different cluster sizes, 
we study the melting-like transition in Rb$_{55}$ and Cs$_{55}$ clusters and
make a comparison with the melting behavior of Na$_{55}$ and K$_{55}$. As the
atomic number increases, the height of the specific heat peaks decreases, 
their width increases, and the
melting temperature decreases as in bulk melting, but in a more pronounced way.
\end{abstract}

\vskip2pc]

\section{Introduction}

The melting transition of a pure bulk material occurs at a well defined
temperature for a given external pressure. The melting-like transition in
small clusters composed of a finite number of atoms spreads instead over a
temperature interval that widens as the cluster size decreases. A number of
thermal properties sensitive to the cluster size and
essentially different from those found in bulk materials
emerge in the transition region defined by that temperature interval, which
has motivated a lot of theoretical 
\cite{Jel86,Bul92,Bla97,Ryt98,Cle98,Cal99,Agu99a,Agu99b} and experimental
\cite{Mar96,Sch97,Sch99,Hab99} investigations. The experiments of Schmidt
{\em et al.}\cite{Sch97} have shown strong nonmonotonic variations
of the melting temperature of free sodium clusters with size, which can not be
completely explained either by electronic or geometric shell closing arguments.
The theoretical simulations have predicted the occurrence of several premelting
effects, like surface melting or structural isomerizations, and also the
existence of a dynamic coexistence regime,
where the cluster can fluctuate in time
between being completely solid or liquid.

We have previously reported density functional
orbital-free molecular dynamics (OFMD)
simulations of the melting process in sodium clusters
Na$_N$, with N=8,20,55,92, and 142.\cite{Agu99a,Agu99b} 
The OFMD technique\cite{Pea93}
is completely analogous to the method devised by Car and Parrinello (CP)
to perform dynamic simulations at an
{\em ab initio} level,\cite{Car85} but the electron density is taken as the
dynamic variable,\cite{Hoh64} as opposed to the Kohn-Sham (KS)
orbitals\cite{Koh65}
in the original CP method. This technique, whose main advantage over KS-based
methods is that the computational effort to update the electronic system scales
linearly with cluster size, has been already used both in
solid state\cite{Sma94,Gov99} and cluster\cite{Bla97,Agu99a,Agu99b,Sha94,Gov95} physics.
Our predictions of the temperatures at which homogeneous cluster melting occurs
were in good agreement with the experiments of Haberland's group,
\cite{Sch97,Sch99} excepting the enhancement of the melting temperature around
N=55, which was not reproduced. We also observed a number of interesting
premelting effects, mostly the establishment of a surface melting stage at a
temperature lower than the homogeneous melting temperature for Na$_{20}$,
Na$_{92}$ and Na$_{142}$, and several isomerization transitions
in Na$_8$ and Na$_{20}$. It is interesting to study similar systems like  
clusters of K, Rb and Cs in order to assess whether those trends are a general 
feature of alkali clusters or not. With this goal, we consider in this paper
the melting-like transition of K$_N$ clusters, with N=20, 55, 92, and 142, and
compare it with that of sodium clusters of the same size. As a second step, for
a fixed cluster size of N=55, we study the melting behavior of Rb$_{55}$ and
Cs$_{55}$ and compare it with that of Na$_{55}$ and K$_{55}$.
In the next section we briefly present some technical details of the method.
The results are presented and discussed in section III and, finally,
section IV summarizes our main conclusions.

\section{Theory}

The orbital-free molecular dynamics method is a Car-Parrinello total
energy scheme\cite{Car85}
which uses an explicit kinetic-energy functional of the 
electron density, and has the electron
density as the dynamical variable, as opposed to the KS single particle
wavefunctions. 
The main features of the energy functional and
the calculation scheme have been described at length in previous work,
\cite{Bla97,Agu99a,Pea93,Sma94,Sha94} and details of our method are as
described by Aguado et al.\cite{Agu99a} In brief, the electronic kinetic 
energy functional of the electron density, $n(\vec r)$, corresponds to 
the gradient expansion around the homogeneous limit through second order 
\cite{Hoh64,Mar83,Yan86,Per92}
\begin{equation}
T_s[n] =
T^{TF}[n] + {\frac{1}{9}} T^W[n],
\end{equation}
where the first term is the Thomas-Fermi functional (Hartree atomic units have
been used)
\begin{equation}
T^{TF}[n] = \frac{3}{10}(3\pi^2)^{2/3}\int n(\vec r)^{5/3}d\vec r,
\end{equation}
and the second is the lowest order gradient correction, where T$^W$,
the von Weizs\"acker term, is given by 
\begin{equation}
T^{W}[n] = 
\frac{1}{8}\int \frac{\mid \nabla n(\vec r) \mid^2}{n(\vec r)}d\vec r.
\end{equation}
The local density approximation is used for exchange and correlation.\cite
{Per81,Cep80} In the external field acting on the electrons,
$V_{ext}(\vec r) = \sum_n v(\vec r -\vec R_n)$, we take $v$ to be
the local pseudopotential of Fiolhais {\em et al.}, \cite{Fio95} which
reproduces well the properties of bulk alkalis and has been shown to have good
transferability to alkali clusters.\cite{Nog96}
The cluster is placed in a unit cell of a cubic superlattice,
and the set of plane waves periodic in the superlattice
is used as a basis set to expand the valence density.
Following Car and Parrinello,\cite{Car85} the coefficients
of that expansion are regarded as generalized coordinates of a set of 
fictitious classical particles, and the corresponding Lagrange equations of 
motion for the ions and the electron density distribution are solved as 
described in Ref. \onlinecite{Agu99a}.

The calculations used a supercell of edge 
71 a.u. for K$_{20}$ and 81 a.u. for K$_{55}$, K$_{92}$, K$_{142}$, Rb$_{55}$
and Cs$_{55}$. An energy cut-off of 8 Ryd was used in the plane wave expansion
of the energy for K clusters, and of 6.15 Ryd for Rb and Cs clusters. 
In all cases, a 64$\times$64$\times$64 grid
was used. The cut-offs used give a
convergence of bond lengths and binding energies as good as that obtained for
sodium clusters.\cite{Agu99a} The fictitious
mass associated with the electron density coefficients
ranged between 6.3$\times$10$^8$ and 4.0$\times$10$^9$ a.u., depending on the
material and on the temperature of the simulations.
The equations of motion of K clusters were integrated using the Verlet 
algorithm\cite{Ver65} for both electrons and ions with a time step 
ranging from $\Delta$t = 0.83 $\times$ 10$^{-15}$ sec. for the simulations 
performed at the lowest temperatures, to $\Delta$t = 0.67 $\times$ 
10$^{-15}$ sec. for those at the highest ones. In the case of Rb$_{55}$ the
time steps ranged from $\Delta$t = 2.38 $\times$ 10$^{-15}$ sec. to
$\Delta$t = 1.31 $\times$ 10$^{-15}$ sec, and in the case of Cs$_{55}$ from
$\Delta$t = 4.29 $\times$ 10$^{-15}$ sec. to $\Delta$t = 1.79 
$\times$ 10$^{-15}$ sec. These choices resulted in 
a conservation of the total energy better than 0.1 \% in all cases. 

Several molecular dynamics simulation runs at different constant energies 
were performed in order to obtain the caloric curve for each cluster.
The initial positions of the atoms for the first run were taken by slightly 
deforming the equilibrium low-temperature geometry of the cluster.
The final configuration of each run served as the starting geometry for the
next run at a different energy. The initial velocities for every new run were
obtained by scaling the final velocities of the preceding run. The total 
simulation times for each run at constant energy were 40 ps for K clusters.
80 ps for Rb$_{55}$, and 140 ps for Cs$_{55}$. These different simulation
times were chosen in order to obtain a good convergence of the several
melting indicators described below. They increase with atomic number because
the typical atomic vibrational frecuencies decrease with atomic number.

A number of indicators to locate the melting-like transition were employed.
Namely, the specific heat defined by \cite{Sug91}
\begin{equation}
C_v = [N - N(1-\frac{2}{3N-6})<E_{kin}>_t ~ <E_{kin}^{-1}>_t]^{-1},
\end{equation} 
where N is the number of atoms and $<>_t$ indicates the average along a
trajectory; the root-mean-square (rms) bond length fluctuation\cite{Jel86} 
\begin{equation}
\delta = \frac{2}{N(N-1)}\sum_{i<j}\frac{(<R_{ij}^2>_t - <R_{ij}>_t^2)^{1/2}}
         {<R_{ij}>_t};
\end{equation}
the ``atomic equivalence indexes''\cite{Bon97}
\begin{equation}
\sigma_i(t) = \sum_j \mid \vec R_i(t) - \vec R_j(t) \mid,
\end{equation}
and finally, the average over a whole dynamical trajectory of the 
radial atomic distribution function g(r), defined by
\begin{equation}
dN_{at} = g(r)dr
\end{equation} 
where $dN_{at}(r)$ is the number of
atoms at distances from the center of mass between r and r + dr.

The experimental results of Haberland {\em et al.}\cite{Sch97,Sch99,Hab99} suggest that both electronic and atomic shell
effects determine the irregular size evolution of the melting temperatures of sodium clusters. Our orbital-free method
does not account for quantum shell effects, and gives cluster energies that vary smoothly as a function of cluster size
(that is, it does not reproduce the energy oscillations associated with electronic shell closures). All atomic shell effects
associated with the geometrical arrangement of ions are properly accounted for, however. Thus, our method is not expected to
give a detailed account of the size variation of the melting temperatures of metal clusters. It will work better for those
cluster sizes where an electronic shell closing appears (N=20, 92 and 142), and worse for those sizes intermediate between
two electronic shell closures (N=55). The material dependent trends for a given cluster size should be accurately predicted
by our energy model, as these will be a consequence of the specific pseudopotential.

\section{Results}

A very important issue in the simulations of cluster
melting is the election of the low-temperature isomer
to be heated. A good knowledge of the ground state structure (global minimum)
is required, as the details of the melting transition are known to be
isomer-dependent.\cite{Bon97} But the problem of performing a realistic
global optimization search is exponentially difficult as size increases, so
finding the global minima of clusters with 55 atoms or more becomes 
impractical. In our previous work\cite{Agu99b} we directly started from 
icosahedral isomers for Na$_{55}$, Na$_{92}$ and Na$_{142}$, 
as there is some experimental\cite{Mar96} and theoretical
\cite{Kum99} indications that suggest icosahedral packing in sodium clusters,
and found a good agreement with the experimental results of Haberland's
group.\cite{Sch97} Simulated annealing runs for Na$_{92}$ and Na$_{142}$ always 
led to disordered structures with an energy higher than that of the 
corresponding icosahedral isomer. The melting behavior of these disordered
structures has been separately analyced and found to be different from that of
icosahedral clusters.\cite{Agu20} As the comparison with experiment was
favourable only for icosahedral isomers and the total energy of these structures
was always lower than that of disordered structures,
we have chosen icosahedral isomers in the
study of the melting behavior of large alkali clusters: K$_{55}$, Rb$_{55}$ and
Cs$_{55}$ are complete two-shell icosahedrons, K$_{92}$ and K$_{142}$ are
incomplete three-shell icosahedrons constructed by following the icosahedral
growing pattern described by Montejano-Carrizales {\em et al.}\cite{Mon96} The
low-temperature isomer of K$_{20}$ was obtained by the dynamic simulated
annealing technique,\cite{Car85} by heating the cluster to 400 K and then
slowly reducing the temperature. The resulting structure is essentially the
same as that obtained for Na$_{20}$ with the same technique.\cite{Agu99a}

The temperature evolutions of the specific heat C$_v$ and of the rms bond length
fluctuation $\delta$ of K$_{20}$ are shown in Fig 1. 
The specific heat displays two maxima
around 90 K and 130 K. The $\delta$(T) curve has a small positive slope at low
temperatures that reflects the thermal expansion of the solidlike cluster, and
then two abrupt increases that correlate with the two peaks in the specific
heat. Both magnitudes indicate that the melting of K$_{20}$ occurs in two
well separated steps over a wide range of temperatures. 
To analyce the nature of those two steps we show in Fig. 2 short-time
averages of the ``atomic
equivalence indexes'' of K$_{20}$ for a number of representative temperatures.
For a temperature at which the cluster is completely solid, the $\sigma_i$(t)
curves show a high degeneracy which is specific of the symmetry of the isomer
under consideration.\cite{Bon97} The structure of
K$_{20}$, as that of Na$_{20}$,\cite{Agu99a} can be divided into two subsets:
two internal ``core'' atoms and 18 peripheral ``surface'' atoms. The transition
at $\approx$ 90 K is identified with an isomerization transition in which the 18
peripheral atoms begin to interchange their positions in the cluster while the
two central atoms remain oscillating around their initial positions. When a
temperature of $\approx$ 130 K is reached, one of the two inner atoms moves out
to the cluster surface, while the other remains in its central position. Then
the second transition is identified with another isomerization transition in 
which a new set of (19+1) isomers begins to be visited. The $\delta$ quantity
increases with temperature after this point in a smooth way. This is due to the
more and more frequent interchanges of the central atom with one of the 
peripheral atoms upon increasing the temperature. Nevertheless,
the interchange rate between
central and peripheral atoms remains slower than the interchange rate between
peripheral atoms for all temperatures considered.

Fig. 3 shows the specific heat and $\delta$ curves for K$_{55}$. The 
specific heat displays a main assymetric peak centered approximately at 160 K,
while $\delta$ shows two abrupt increases at $\approx$ 110 K and 160 K.
The second abrupt increase in $\delta$ coincides with the position of the main
specific heat peak. Although the first step in $\delta$ is not in correspondence
with any well-defined specific heat peak, there is a visible shoulder in the
low temperature side (a clear assymetry) of that peak. Moreover, the width of
the transition region, approximately 100 K, is predicted to be the same with 
both indicators. The nature of melting is analyced by plotting the temperature
evolution of the time-averaged radial atomic density distribution g(r). At a
low temperature of T=86 K, Fig. 4 shows that the atoms are distributed in
several icosahedral shells (in the outer shell, the twelve atoms in vertex
positions can be distinguished from the rest due to slightly different radial
distances). At the temperature where the first step in $\delta$ emerges, the
detailed structure in the ionic density distribution has been washed out by 
the thermal effects, and the
movies show that the cluster surface is melted. However,
the different shells are still clearly distinguished, showing 
that there are not interchanges between atoms in different shells. At a
temperature higher than 160 K, the disctintion of the several radial shells is
not possible anymore. All the atoms are able to diffuse across the cluster
volume, that is both intrashell and intershell displacements are allowed, and
the liquid phase is completely established. Upon a further increasing in 
temperature, the only appreciable change in g(r) is due to the thermal
expansion of the cluster.

The results for K$_{92}$ are shown in Figs. 5 and 6. Both the specific heat
and $\delta$ predict a two-step melting process, with a first transition at
$\approx$ 110 K and a second transition at $\approx$ 200 K. As seen in Fig. 6,
the first transition
is associated again with surface melting, with no substantial intershell
diffusion. For temperatures higher than 200 K, the cluster is 
completely liquid. K$_{142}$ melts in two main steps, at $\approx$ 140 K and
230 K (Fig. 7). There is also
a small bump on the low-temperature side of the first
specific heat peak, correlated with a small abrupt increase of $\delta$
at $\approx$ 90 K. This previous step is associated with an isomerization
regime in which different isomers preserving the icosahedral symmetry are
visited, and was also found for Na$_{142}$.\cite{Agu99b} These isomerizations
involve the motion of the five vacancies in the outer shell. The surface
melting stage is not developed yet, however, as the icosahedral order persists.
The distribution of atoms in three shells is still distinguished
at a temperature of 190 K where the cluster surface is melted. 
The average radial ionic density 
distribution is not completely uniform until the homogeneous melting temperature
is reached.

The bulk melting temperature of K (337 K)\cite{Ash76}
is reduced by a 10 \% with respect to that of Na (371 K).\cite{Ash76}
The melting temperatures of K clusters are also smaller than those of Na 
clusters\cite{Agu99a,Agu99b}
for all the sizes studied. The percentage reduction in melting temperature is
substantially larger than in the bulk and a slightly decreasing
function of cluster size (19 \%, 17 \%, 16 \% and 15 \% for
N=20, 55, 92, and 142, respectively). The nature of the several premelting
effects observed are similar for both materials. Nevertheless, they are more
easily established in the case of K clusters. For example, surface and
homogeneous melting temperatures were closer together in the case of Na$_{142}$
(240 K and 270 K, respectively)\cite{Agu99b}
than they are for K$_{142}$ (140 K and 230 K,
respectively); while two-step melting was not observed for a perfect
two-shell Na$_{55}$ icosahedron,\cite{Agu99b} the melting surface stage is
well established for K$_{55}$, which has the same low-temperature structure;
in general, the transition region is wider for K than for Na clusters.

As the main points in this comparison are quite independent of cluster size,
we analyce in the following the melting behavior of Rb and Cs clusters for a
fixed cluster size, namely N=55. Specific heat and $\delta$ curves as a function
of temperature are given in Figs. 9 and 10. The results are similar to those
obtained for K$_{55}$, namely a main assymetric specific heat peak and two
steps in $\delta$, the last of which correlates with the peak in the specific 
heat. The radial ionic density distributions of both materials present a
similar temperature dependence and only the results for Cs$_{55}$ are shown in
Fig. 11. The first transition (at $\approx$ 110 K for Rb$_{55}$ and $\approx$
90 K for Cs$_{55}$) is identified with surface melting: Fig. 11 at 110 K
shows that
intershell diffusion is not important yet. The second (at $\approx$ 140 K for
Rb$_{55}$ and 130 K for Cs$_{55}$ ) corresponds to homogeneous melting. We
find that the different alkali clusters with N=55 atoms melt in a similar way. 
The main
differences are the following: a) The homogeneous melting temperature decreases
with increasing atomic number as in the bulk case, but in a more pronounced
way. Specifically, in the series Na$\rightarrow$K$\rightarrow$Rb$\rightarrow$Cs,
the bulk melting temperatures decrease by percentage values of 10 \%, 7 \%, and
3 \%, respectively,\cite{Ash76} while for the 55-atom clusters the corresponding
percentage values are 17 \%, 12.5 \%, and 7 \%, respectively; 
b) The height of the
specific heat peaks decreases and their width increases with increasing atomic
number; c) Premelting effects are more important for the heavier elements.
Specifically, two-step melting was not observed in the case of Na$_{55}$,
\cite{Agu99b} while
a well defined surface melting stage is observed in the thermal evolution of 
K$_{55}$, Rb$_{55}$ and Cs$_{55}$. 

It is perhaps not surprising that the melting temperature reduction is larger in
clusters compared to the bulk phase, where coordination effects associated with
a large proportion of atoms in surface-like positions do not appear. 
But a meaningful comparison
can not be done due to the different structures adopted by alkali elements in
the cluster (icosahedral packing) and bulk (bcc packing) phases. The other two
points do not invoke any comparison with the bulk phase, and can be more
conveniently addressed. Rey {\em at al.}\cite{Rey98} have analyced the influence
of the softness of the repulsive core interaction on cluster melting. 
Specifically, a series of pair potentials differing just in their shape in the
core region was constructed and used to investigate the melting behavior of
13-particle clusters. For those potentials with soft core repulsion, two-step
melting was observed: the first step corresponds to the onset of frequent
isomerizations involving only the surface atoms, while the second corresponds
to homogeneous melting, involving also the central atom. For the harder
potentials, those two steps merge into one, and melting-in-steps processes do
not appear. The repulsive part of our pseudopotential is harder the lighter is
the alkali element,\cite{Fio95} so the importance of premelting effects can be
expected to increase in the series 
Na$\rightarrow$K$\rightarrow$Rb$\rightarrow$Cs. In effect,
a well-defined surface melting stage is not observed for Na$_{55}$, while it
develops before the homogeneous melting point for the heavier alkali elements.
Moseler and Nordiek\cite{Mos99} have studied the influence of the potential
range on the heat capacity of 13-atom Morse clusters. They have found that
decreasing the range of the potential increases the peak of the heat capacity
in the melting transition region. The range of our pseudopotential increases
with atomic number for the alkali elements.\cite{Fio95} We have performed a series of static calculations for
the K$_2$, Rb$_2$ and Cs$_2$ molecules in order to construct their binding energy curves, and have found that the range of
the interatomic interaction increases together with the range of the pseudopotentials,
so a decrease in the
height of the specific heat peak is expected in the series
Na$\rightarrow$K$\rightarrow$Rb$\rightarrow$Cs. This is what is observed indeed.
Thus, we conclude that melting proceeds in a qualitatively similar way in
clusters of the alkali elements Na, K, Rb and Cs, and that the small existing
differences can be explained in terms of the different parameters defining the
corresponding local pseudopotentials.

A few comments regarding the quality of the simulations and of the annealing
runs are perhaps in order here.
The orbital-free representation of the atomic interactions is much more
efficient than the more accurate KS treatments, but is still
substantially more expensive
computationally than a simulation using
phenomenological many-body potentials. Such potentials contain
a number of parameters that are usually chosen by fitting some bulk and/or
molecular properties. In contrast
our model is free of external parameters, although there are
approximations in the kinetic and
exchange-correlation functionals. 
The orbital-free scheme accounts, albeit approximately, for the effects of the
detailed electronic distribution on the total energy and the forces on the ions.
We feel that this is particularly important in metallic clusters for which a
large proportion of atoms are on the surface and experience a very different
electronic environment than an atom in the interior. Furthermore, the
adjustment of the electronic structure and consequently the energy and forces
to rearrangements of the ions is also taken into account.
But the price to be paid
for the more accurate description of the interactions is a less complete
statistical sampling of the phase space. The simulation times are substantially
shorter than those that can be achieved in phenomenological simulations.
Nevertheless, we expect that the
locations of the several transitions are reliable, because 
all the indicators we have
used, both thermal and structural ones, are in essential agreement on
the temperature at which the transitions start.

\section{Summary}

The melting-like transition in K$_N$, with N=20, 55, 92 and 142, Rb$_{55}$
and Cs$_{55}$ clusters has been 
investigated by applying an orbital-free, density-functional molecular
dynamics method. The computational effort which is required is modest in 
comparison with the traditional Car-Parrinello Molecular Dynamics technique 
based on Kohn-Sham orbitals, that would be very costly for clusters of these
sizes. The details of the several transitions have been explained and found to
be similar to those found in the melting-like transition of sodium clusters:
\cite{Agu99a,Agu99b} alkali clusters show generally
a separate surface melting stage prior to homogeneous melting. The homogeneous
melting temperature has been found to decrease with increasing atomic number as
in the bulk limit, but the percentage value of this temperature reduction is
larger than for the bulk materials, due to the larger proportion of surface-like
atoms existing in the cluster; in fact, the reduction in melting temperature
when passing from Na to K clusters slightly decreases with increasing cluster
size. The height of the specific heat peaks decreases and their width increases
with increasing atomic number, and the premelting effects are more important for
the heavier alkalis. These trends have been rationalized in terms of physical
features of the local pseudopotentials employed.

$\;$

$\;$

$\;$

{\bf ACKNOWLEDGMENTS:} 
This work has been supported by DGES 
(Grant PB98-0368)
and Junta de Castilla y Le\'on (VA70/99).
The author is indebted to M. J. Stott, who developed the program which formed the basis of this work. He
also acknowledges useful discussions with J. M. L\'opez and J. A. Alonso.


{\bf Captions of Figures.}

{\bf Figure 1} Specific heat (a) and $\delta$ (b) curves
of K$_{20}$, taking the
internal cluster temperature as the independent variable. 
The deviation around the mean
temperature is smaller than the size of the circles.

{\bf Figure 2} Short-time averaged distances $<r_i(t)>_{sta}$ between each atom
and the center of mass in K$_{20}$, as functions of time for 
four representative values of the internal temperature.
The bold lines follow the evolution of particular atoms.

{\bf Figure 3} Specific heat (a) and $\delta$ (b) curves
of K$_{55}$, taking the
internal cluster temperature as the independent variable. 
The deviation around the mean
temperature is smaller than the size of the circles.

{\bf Figure 4} Time
averaged radial atomic density distribution of
K$_{55}$, at four representative temperatures.

{\bf Figure 5} Specific heat (a) and $\delta$ (b) curves
of K$_{92}$, taking the
internal cluster temperature as the independent variable. 
The deviation around the mean
temperature is smaller than the size of the circles.

{\bf Figure 6} Time
averaged radial atomic density distribution of
K$_{92}$, at four representative temperatures.

{\bf Figure 7} Specific heat (a) and $\delta$ (b) curves
of K$_{142}$, taking the
internal cluster temperature as the independent variable. 
The deviation around the mean
temperature is smaller than the size of the circles.

{\bf Figure 8} Time
averaged radial atomic density distribution of
K$_{142}$, at four representative temperatures.

{\bf Figure 9} Specific heat (a) and $\delta$ (b) curves
of Rb$_{55}$, taking the
internal cluster temperature as the independent variable. 
The deviation around the mean
temperature is smaller than the size of the circles.

{\bf Figure 10} Specific heat (a) and $\delta$ (b) curves
of Cs$_{55}$, taking the
internal cluster temperature as the independent variable. 
The deviation around the mean
temperature is smaller than the size of the circles.

{\bf Figure 11} Time
averaged radial atomic density distribution of
Cs$_{55}$, at four representative temperatures.

\onecolumn[\hsize\textwidth\columnwidth\hsize\csname
@onecolumnfalse\endcsname

\begin{figure}
\psfig{figure=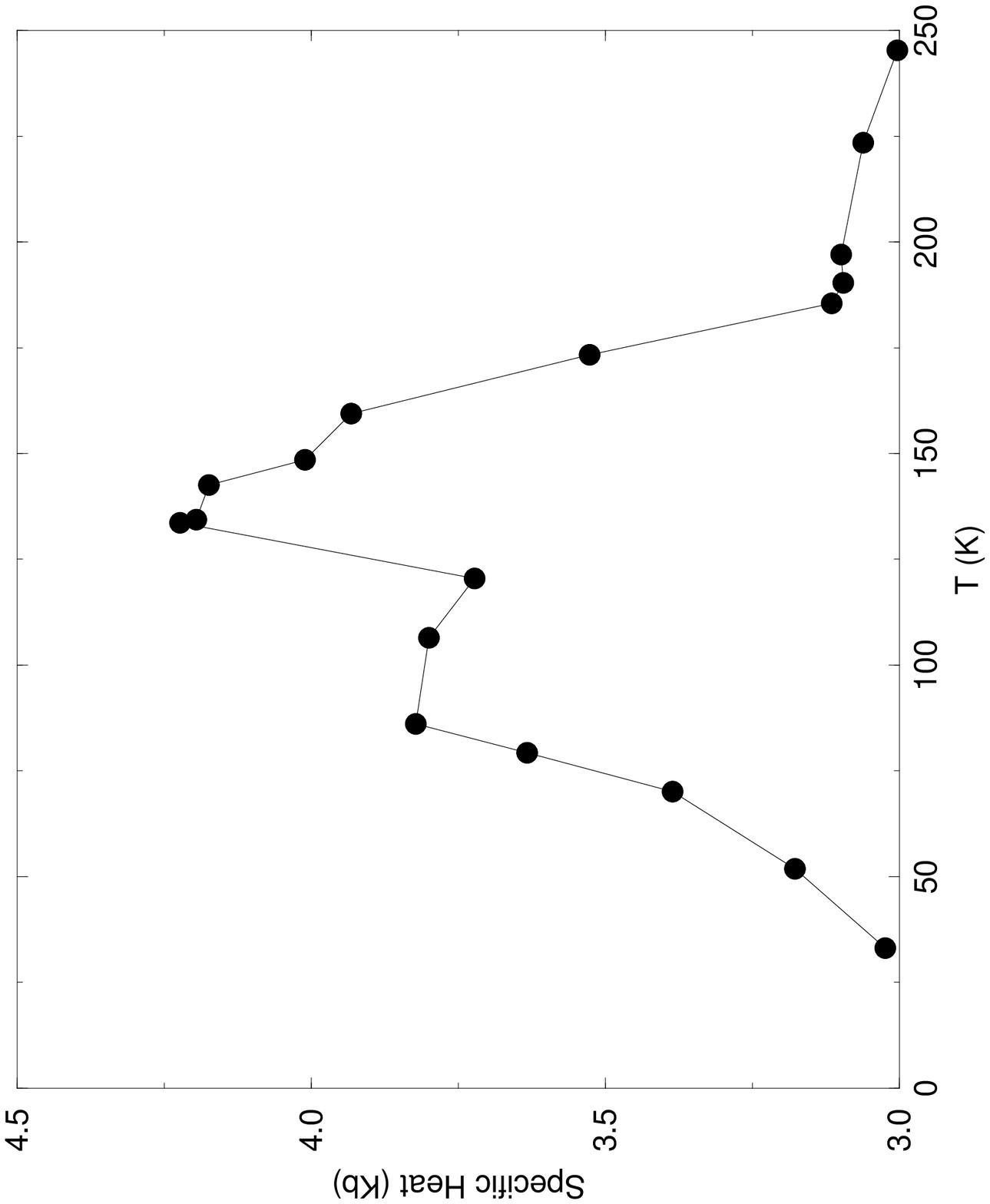}
\end{figure}

\begin{figure}
\psfig{figure=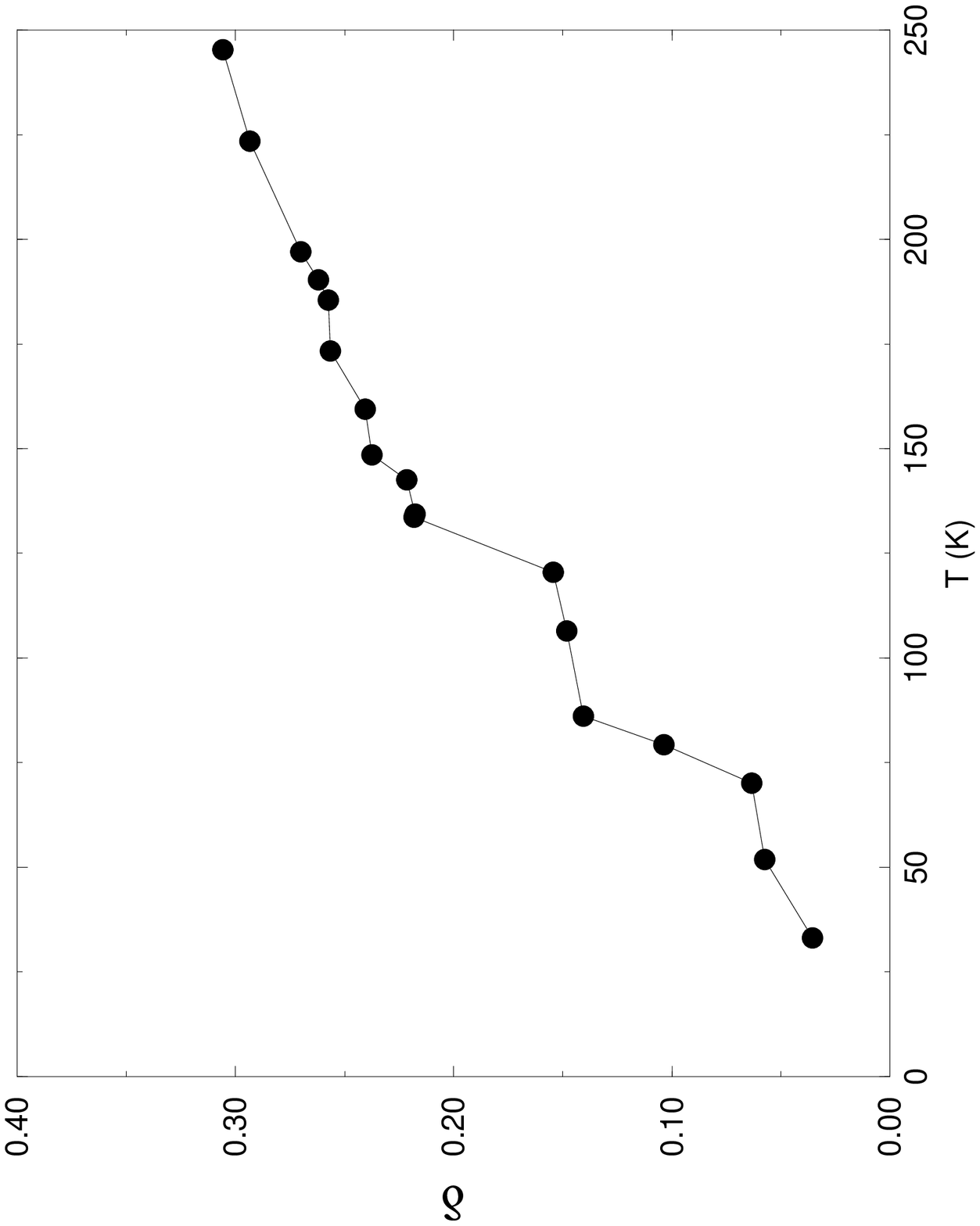}
\end{figure}

\begin{figure}
\psfig{figure=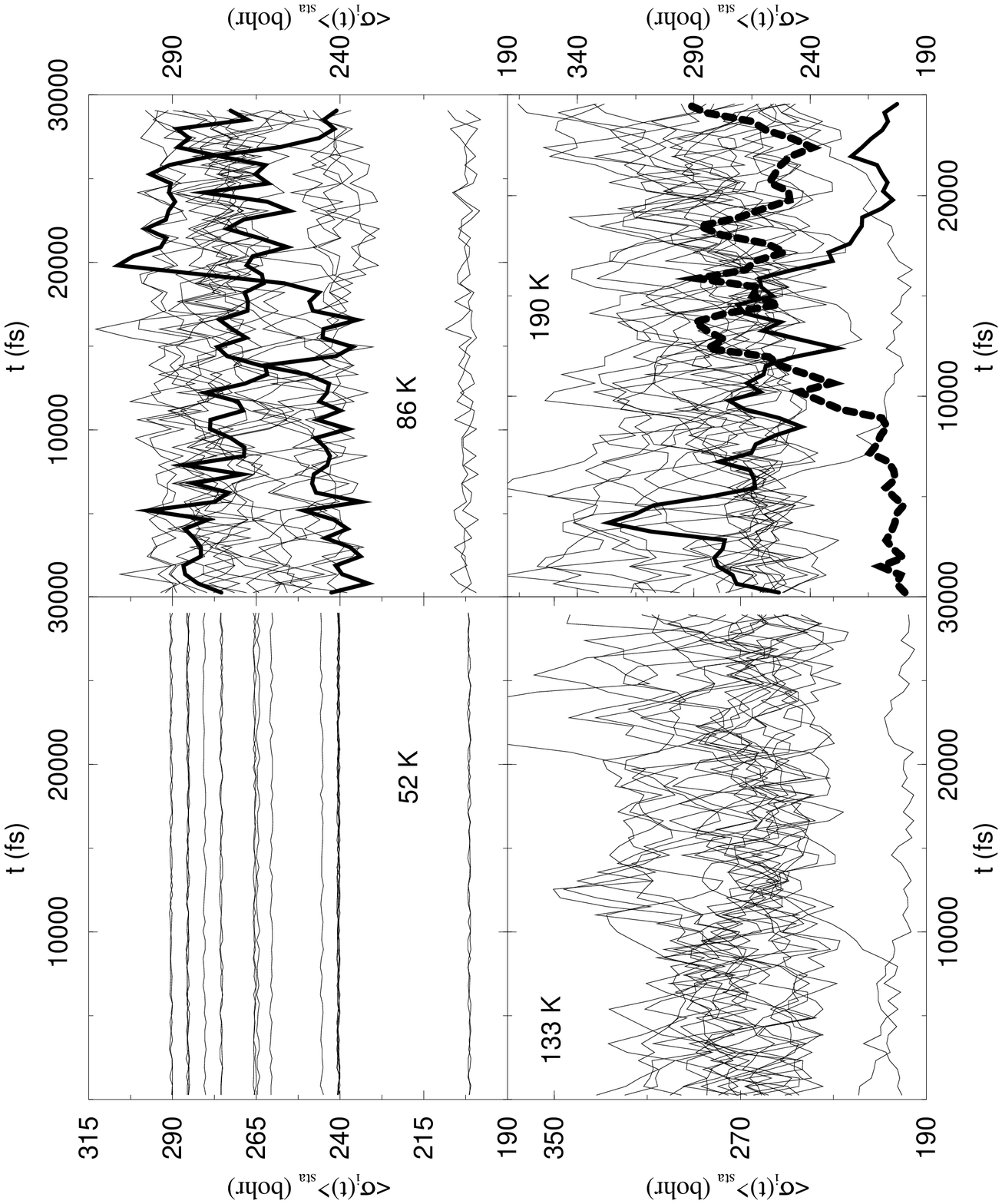}
\end{figure}

\begin{figure}
\psfig{figure=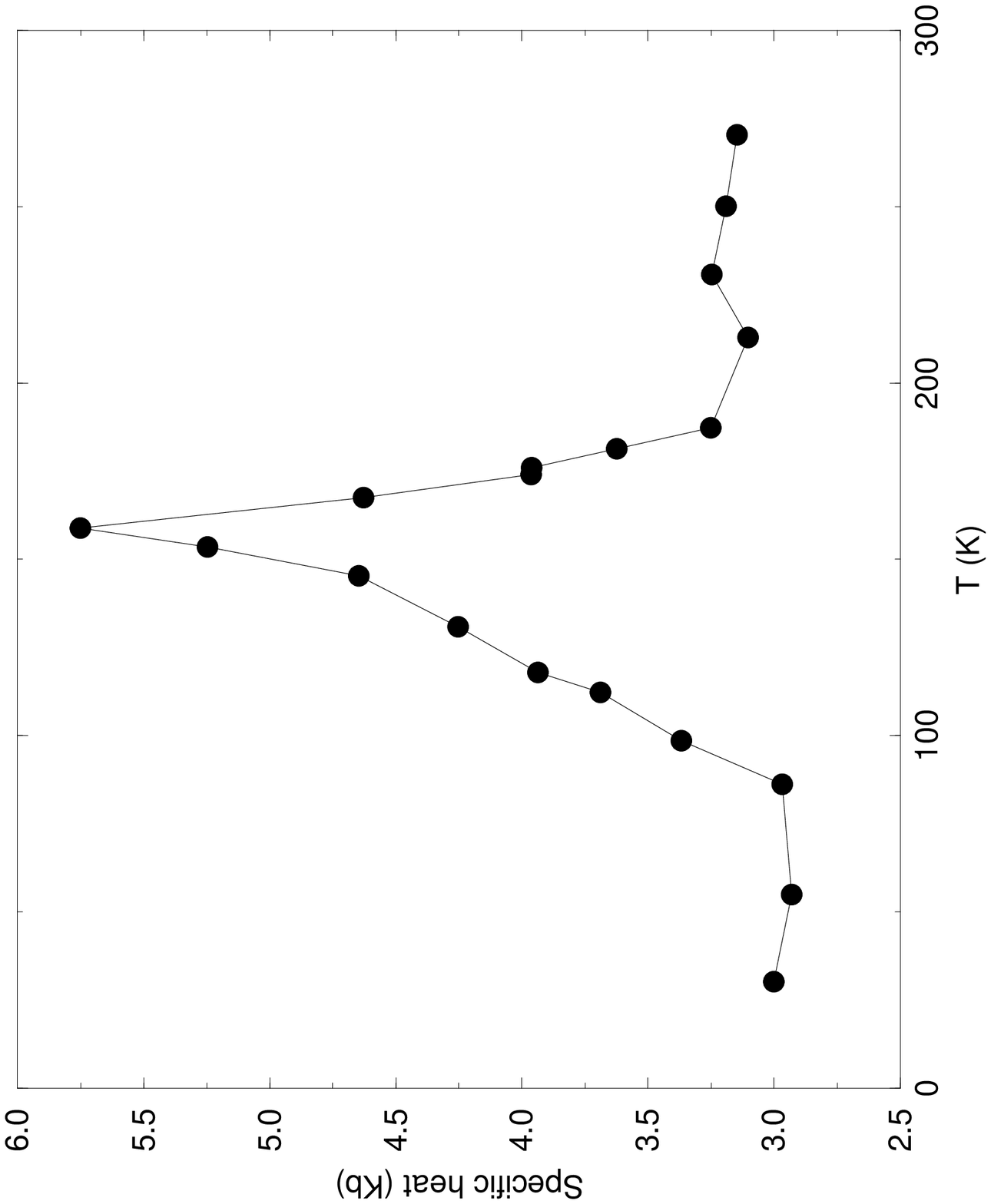}
\end{figure}

\begin{figure}
\psfig{figure=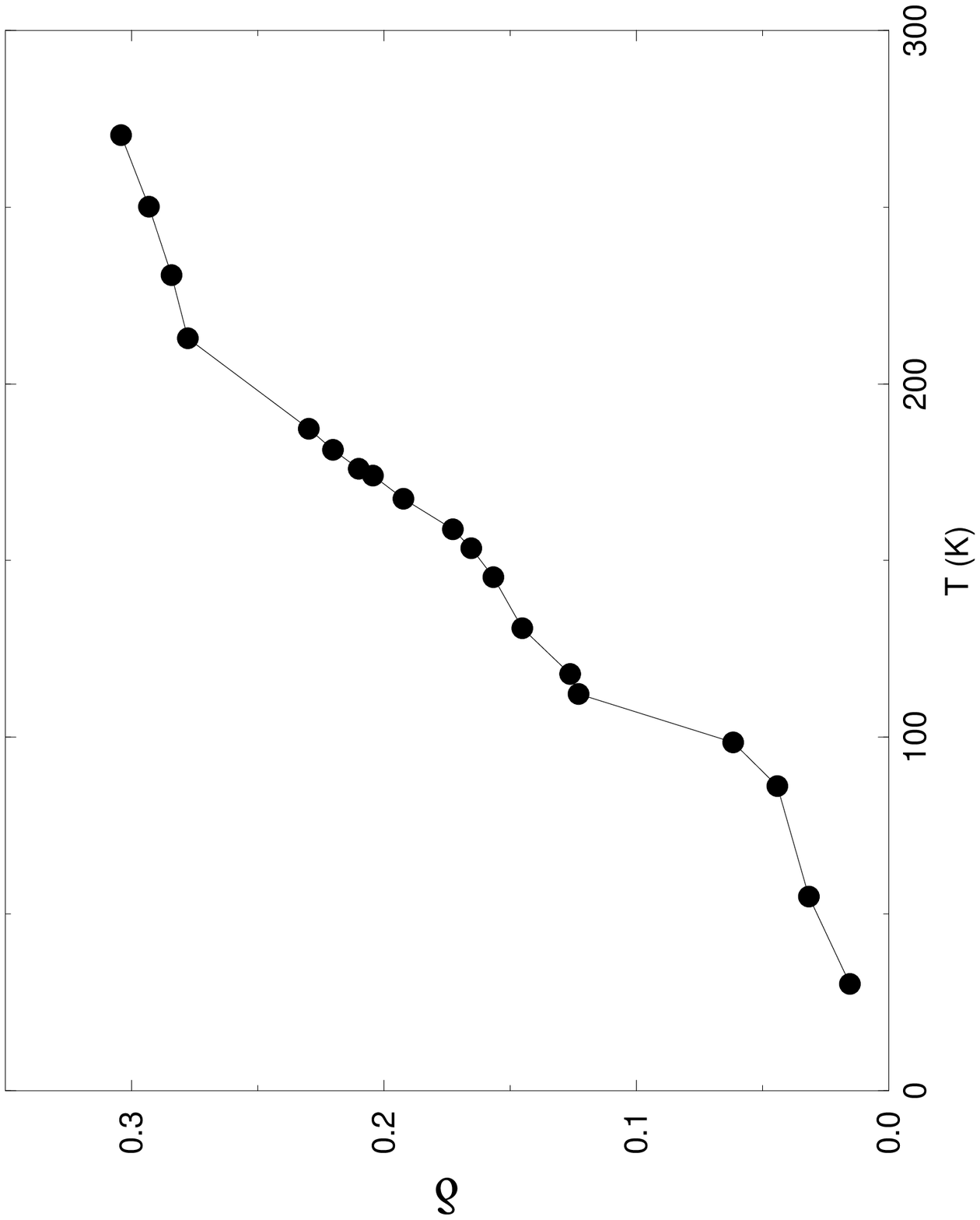}
\end{figure}

\begin{figure}
\psfig{figure=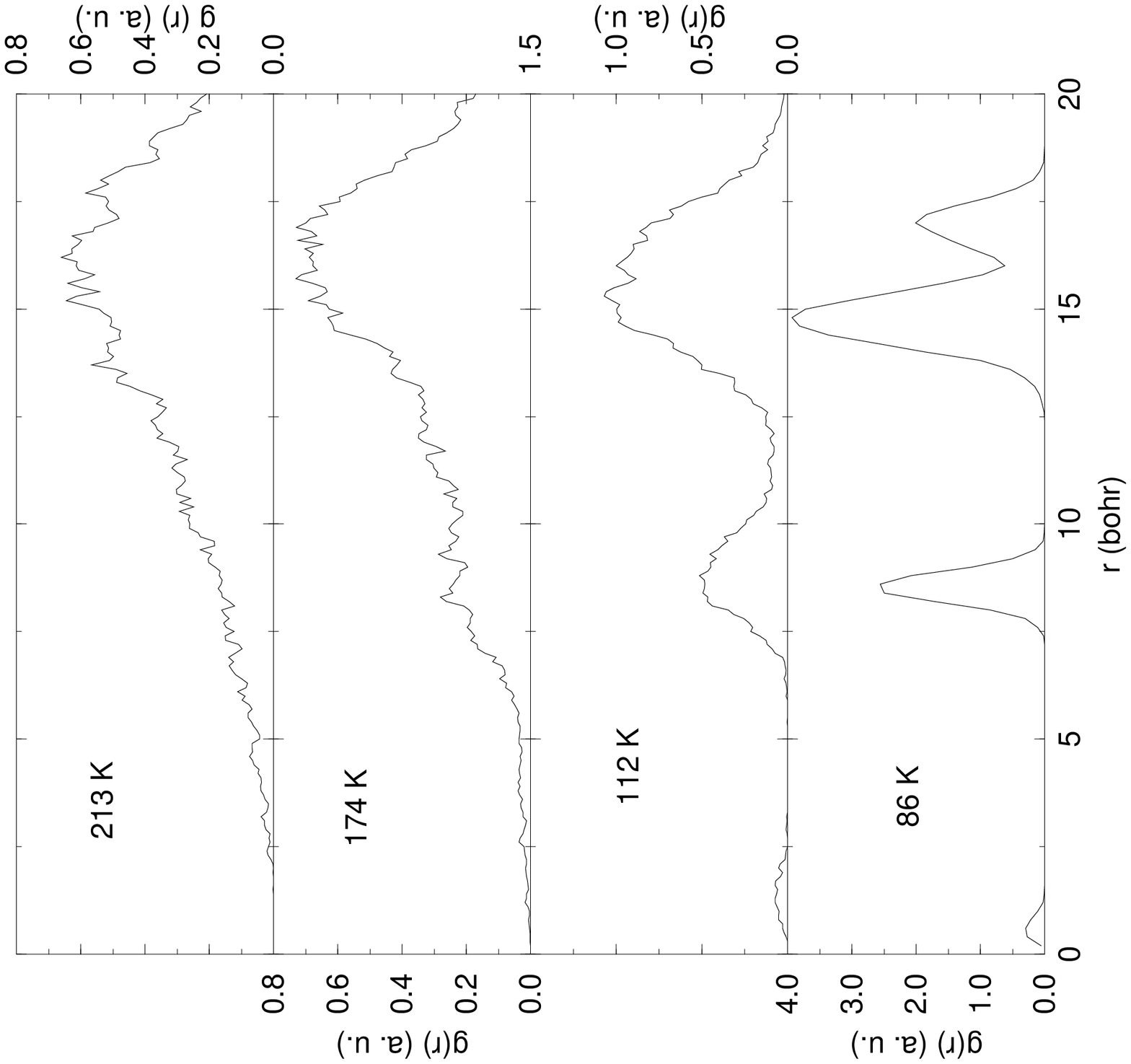}
\end{figure}

\begin{figure}
\psfig{figure=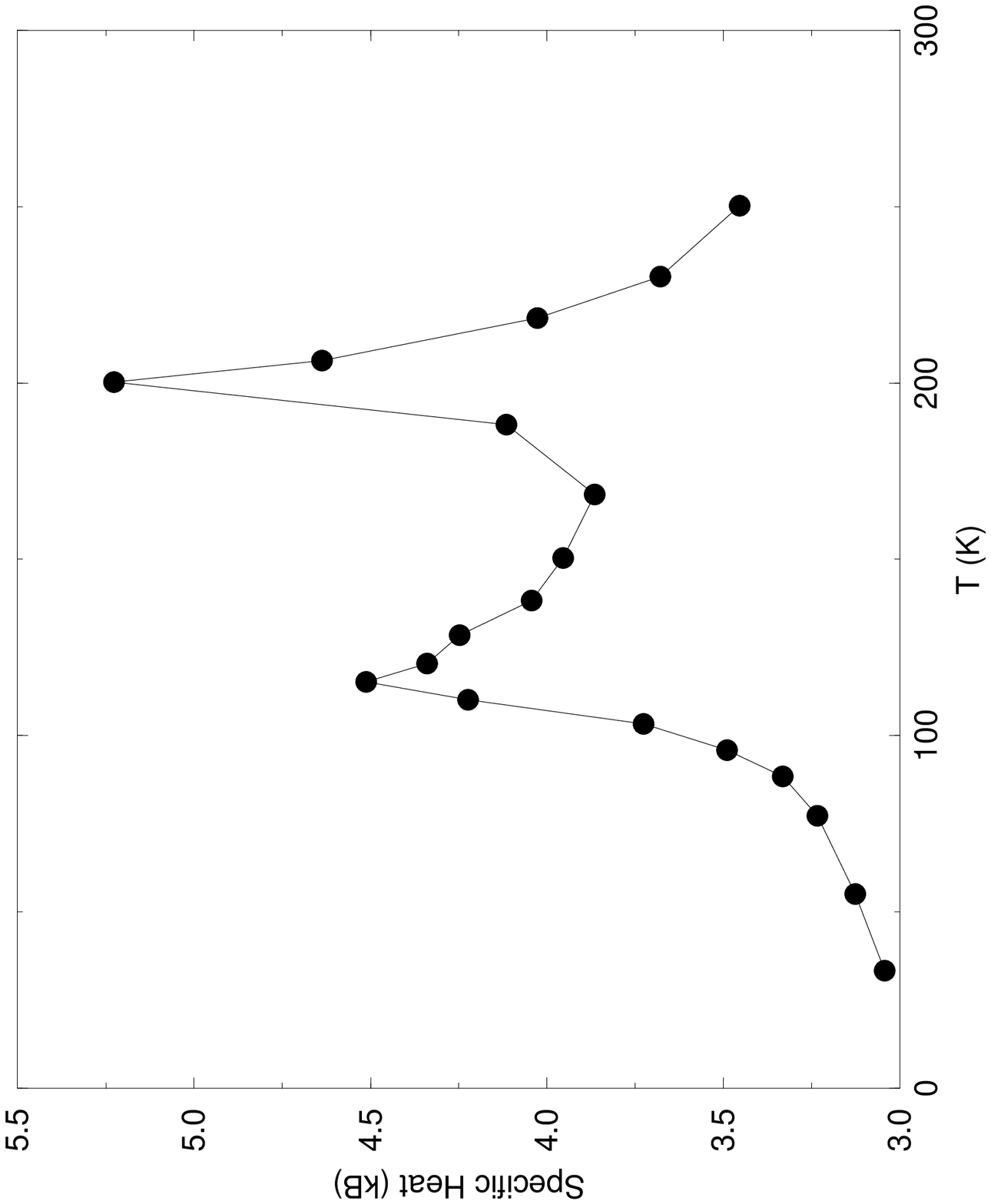}
\end{figure}

\begin{figure}
\psfig{figure=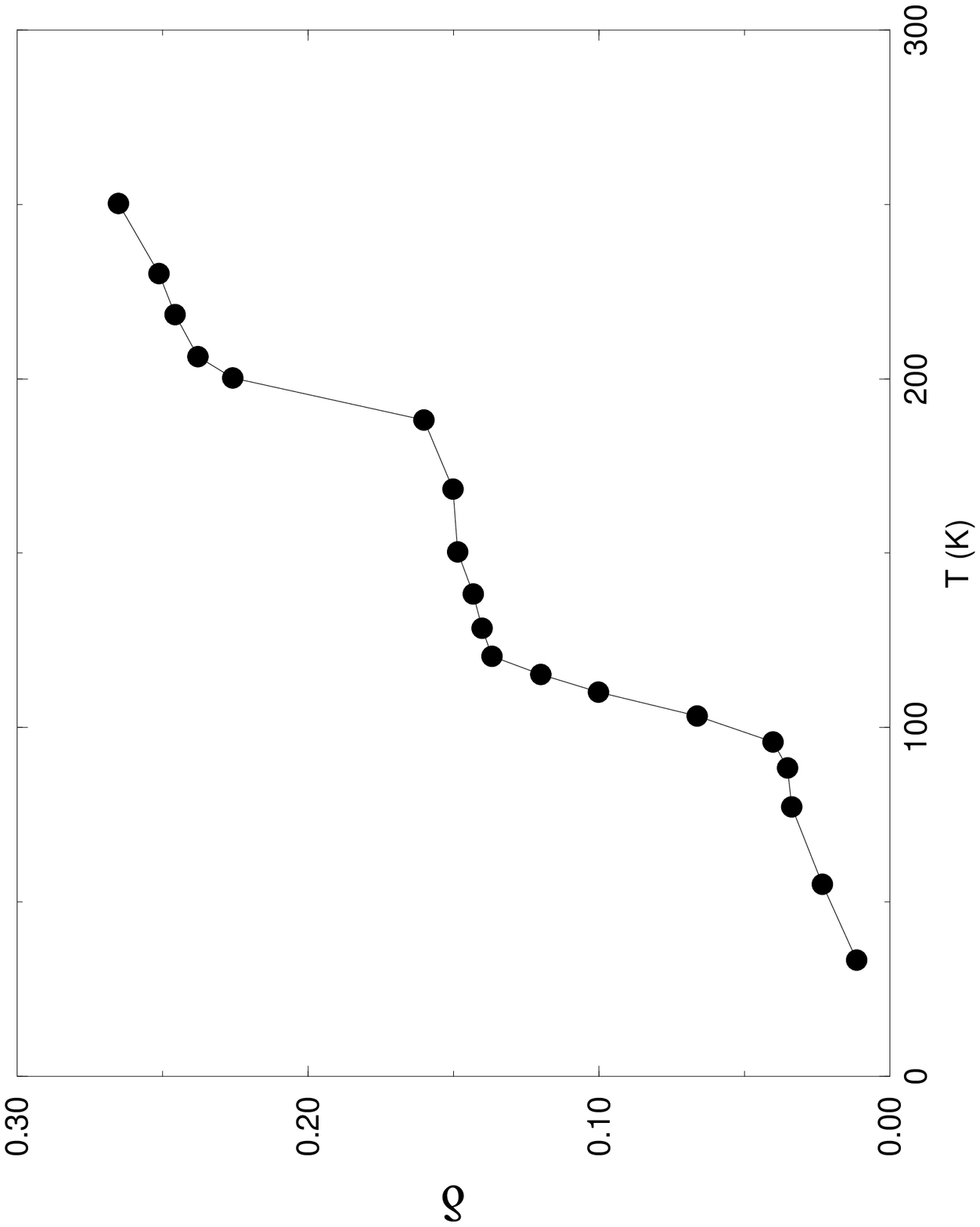}
\end{figure}

\begin{figure}
\psfig{figure=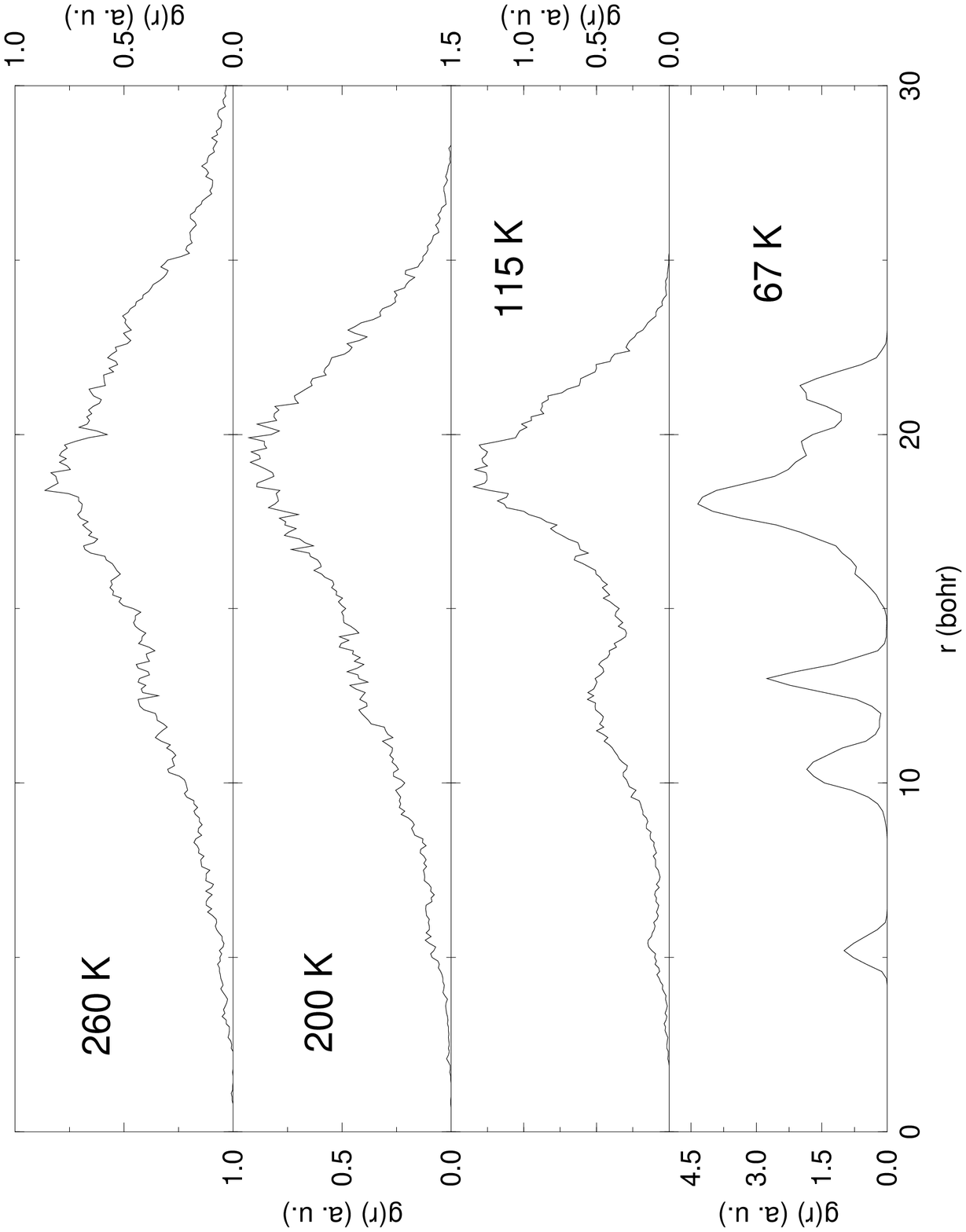}
\end{figure}

\begin{figure}
\psfig{figure=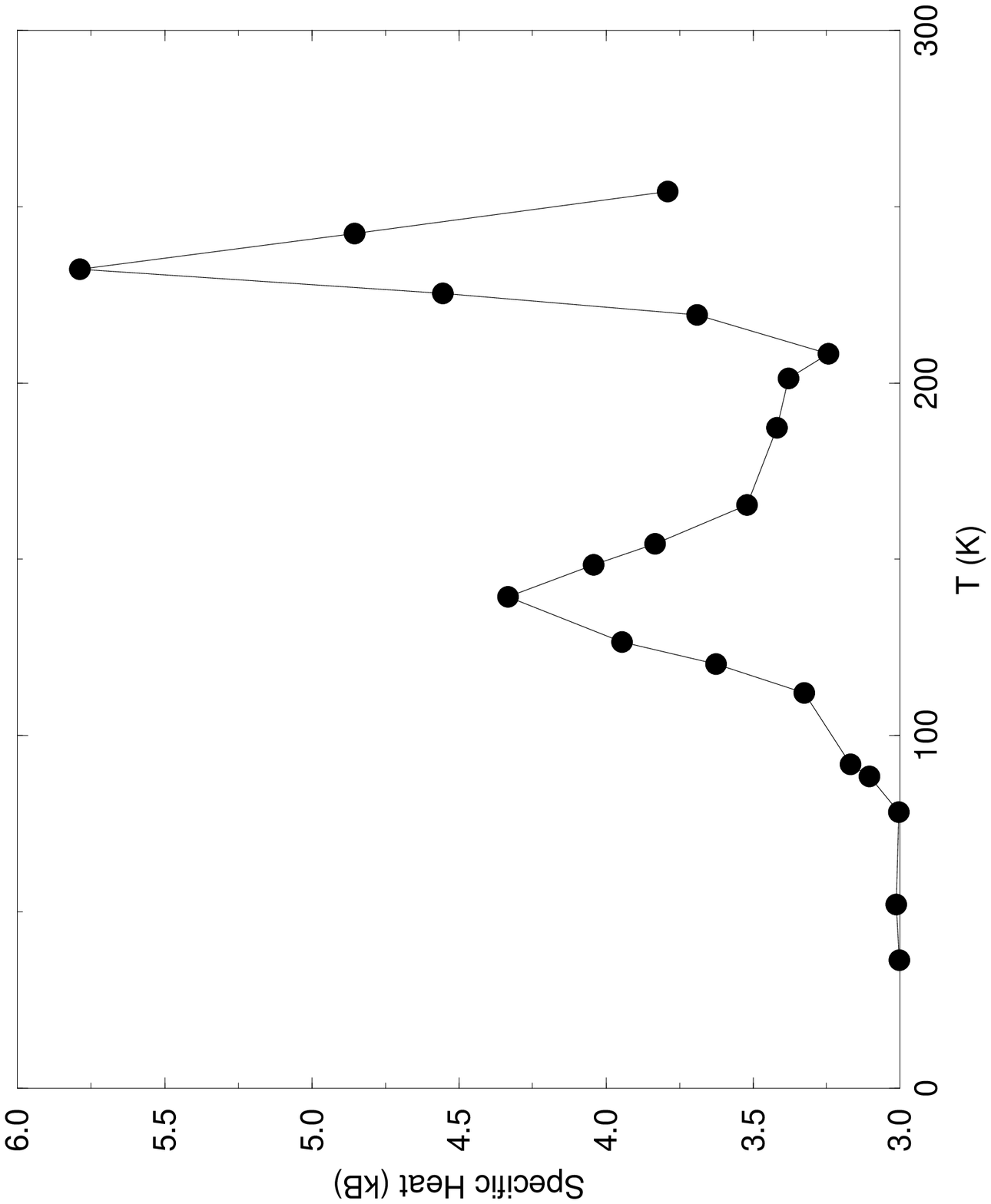}
\end{figure}

\begin{figure}
\psfig{figure=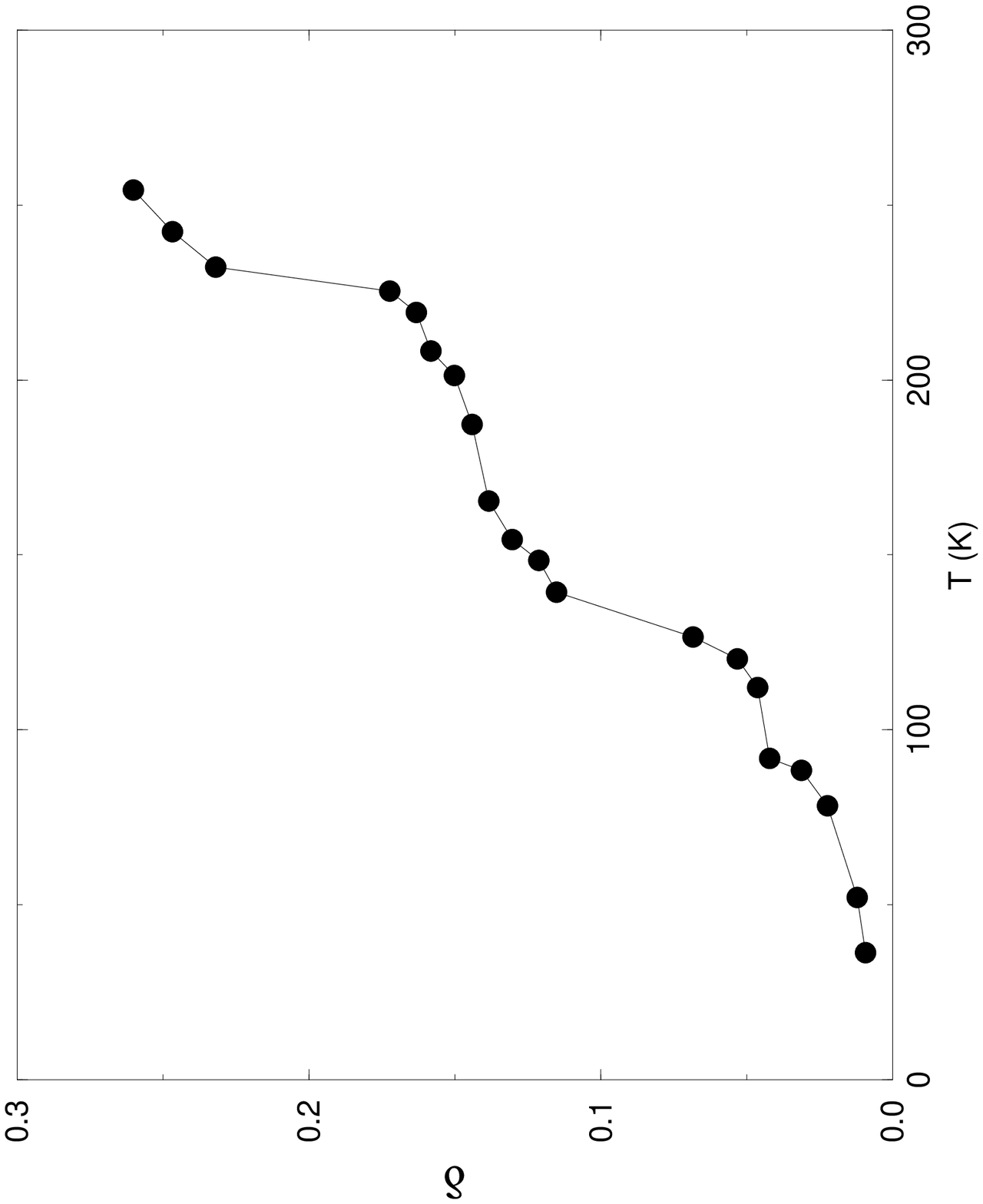}
\end{figure}

\begin{figure}
\psfig{figure=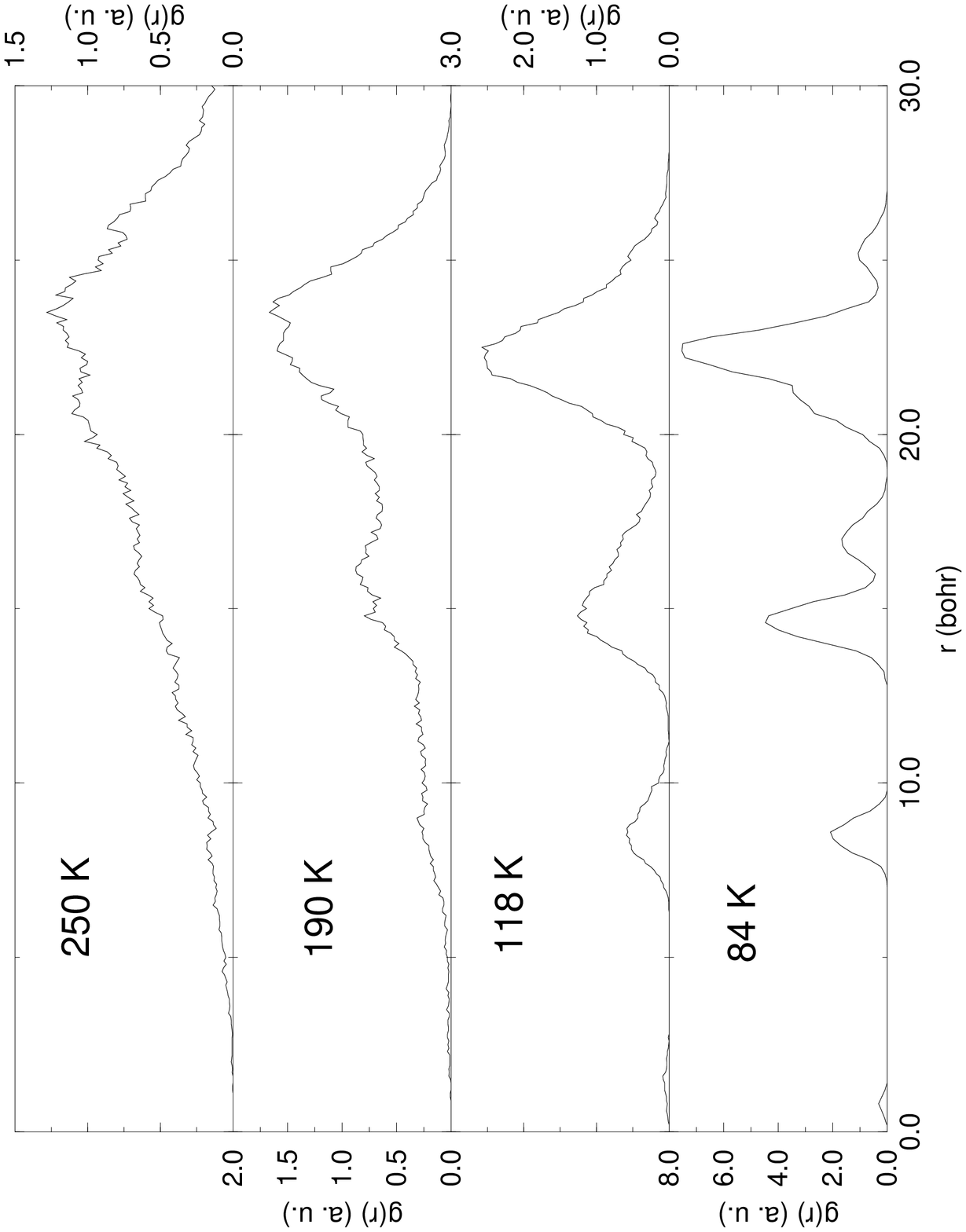}
\end{figure}

\begin{figure}
\psfig{figure=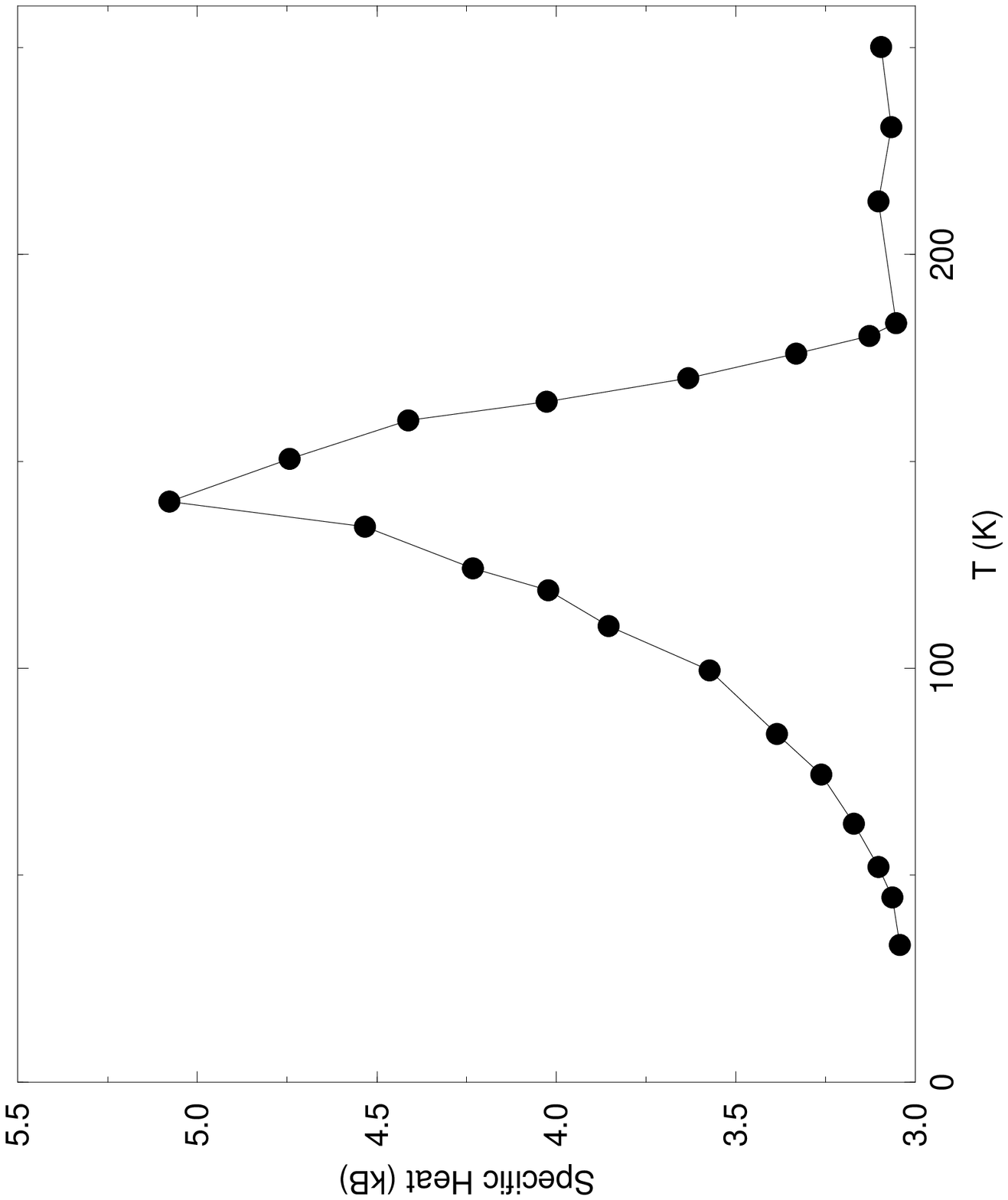}
\end{figure}

\begin{figure}
\psfig{figure=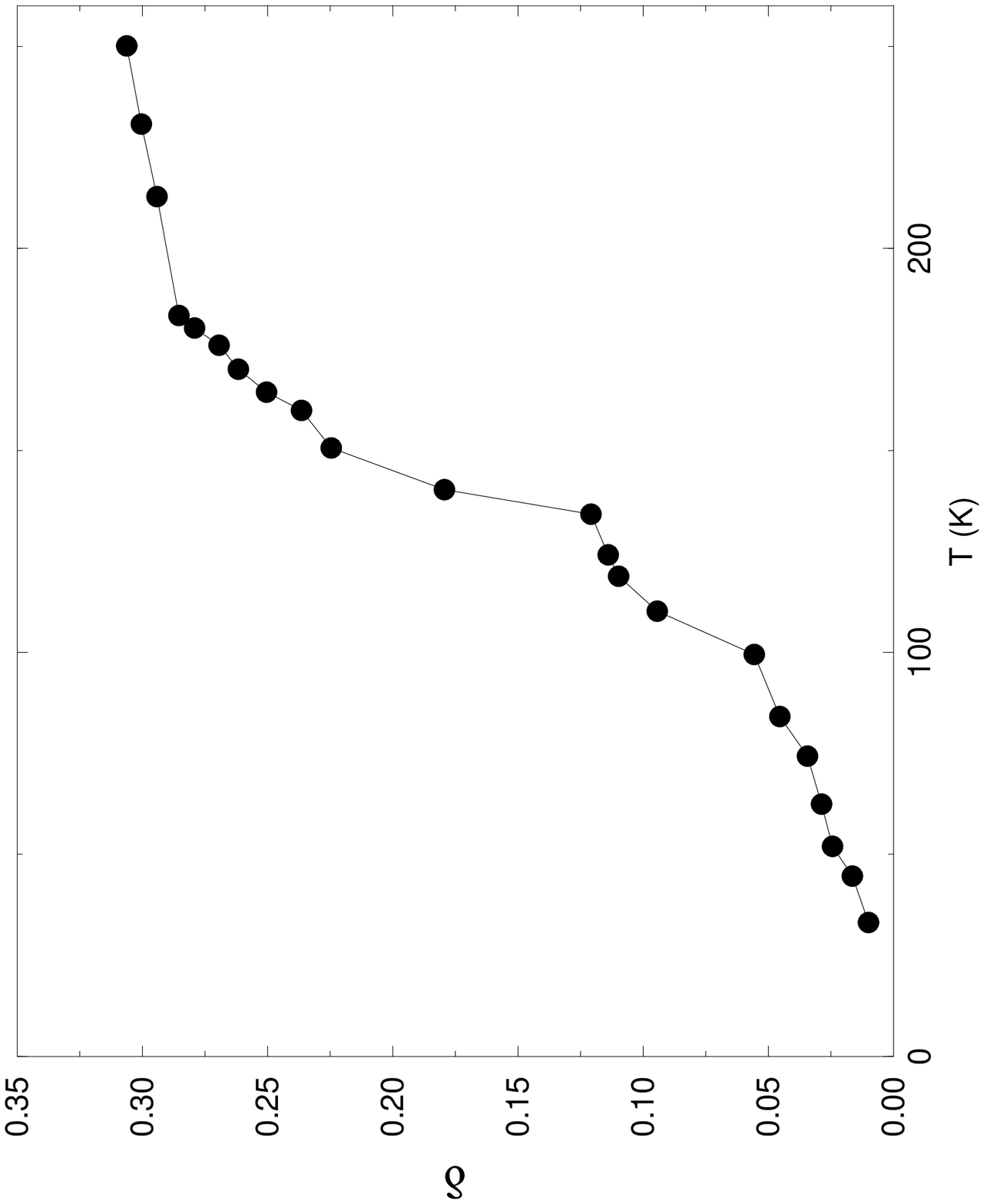}
\end{figure}

\begin{figure}
\psfig{figure=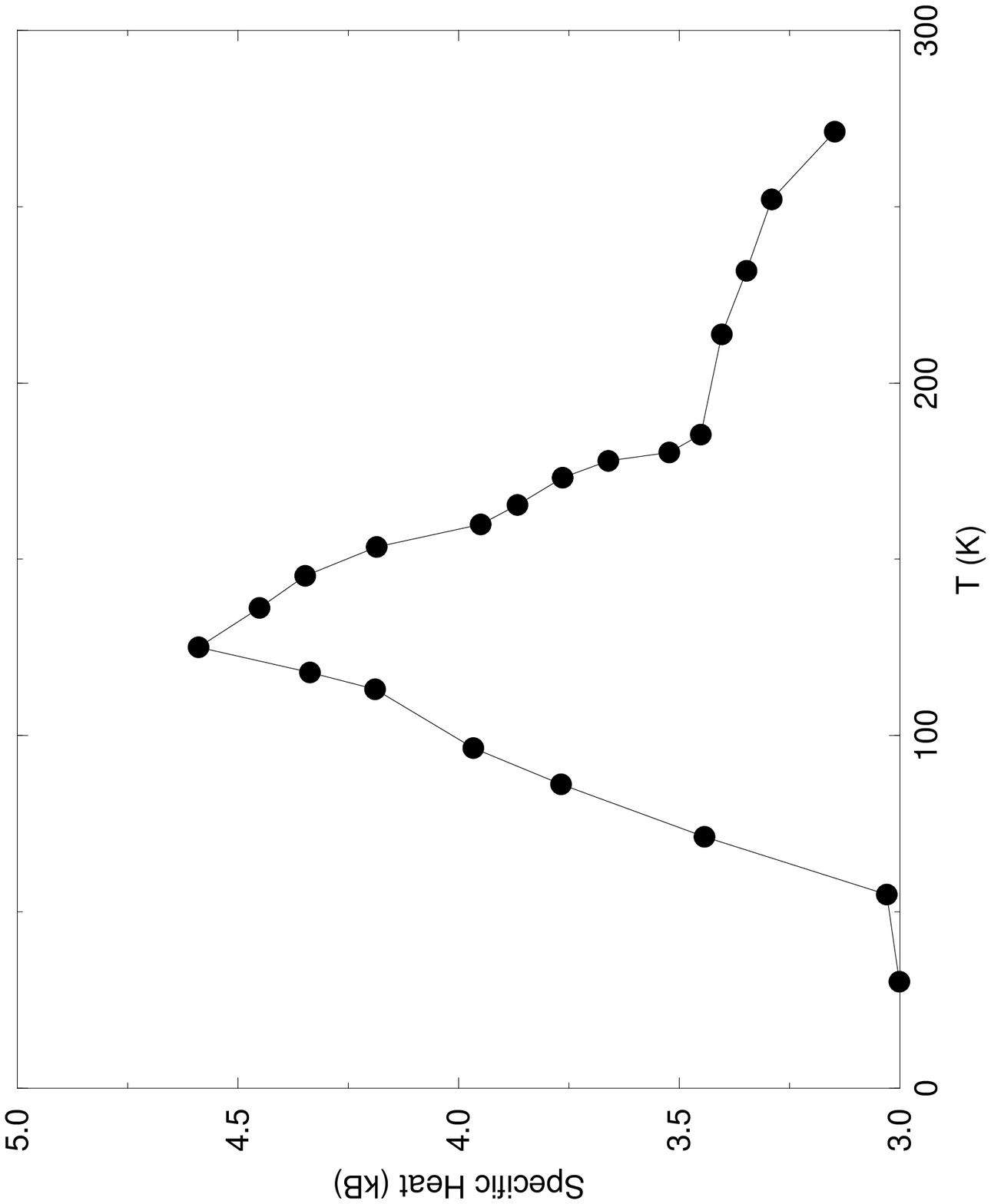}
\end{figure}

\begin{figure}
\psfig{figure=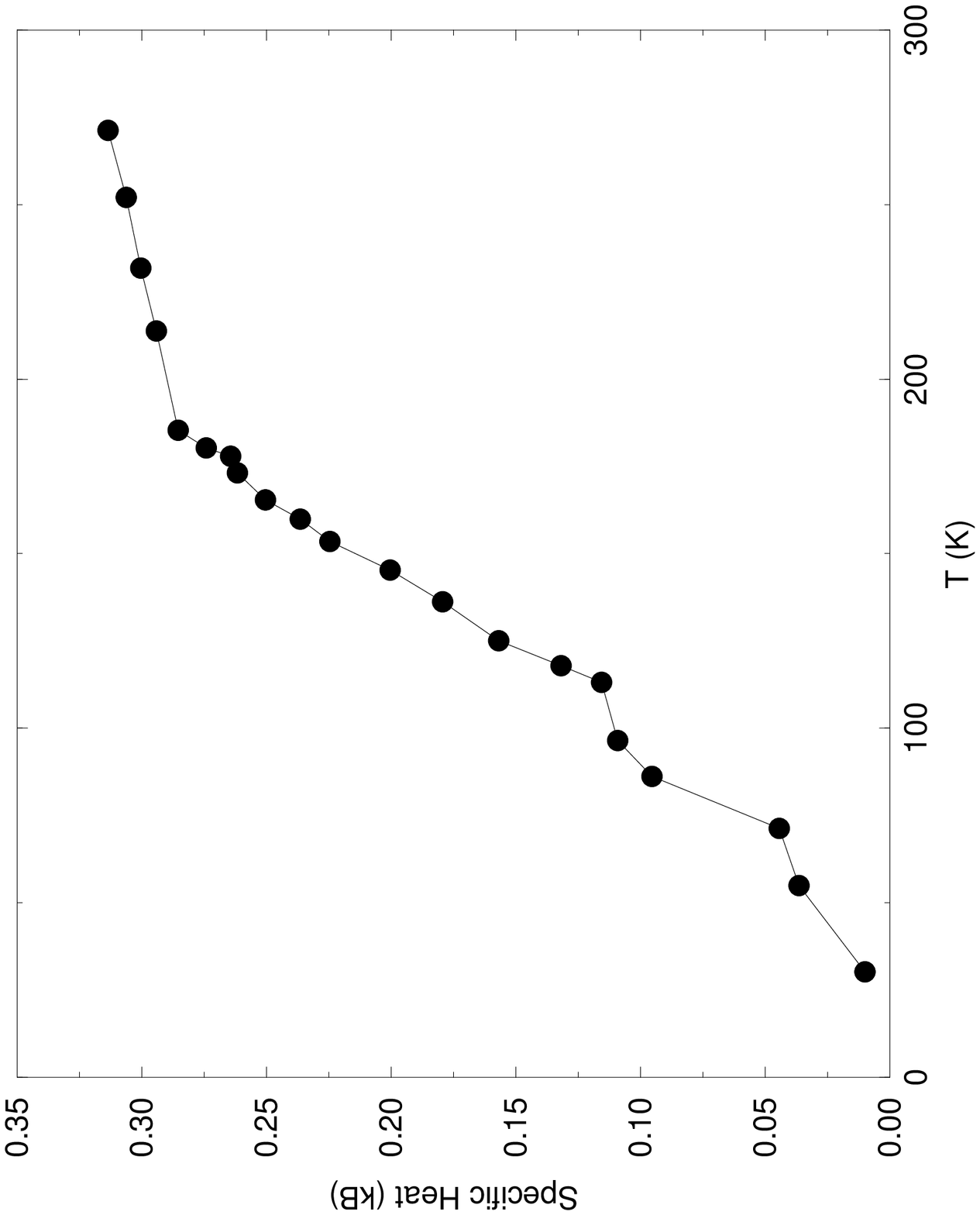}
\end{figure}

\begin{figure}
\psfig{figure=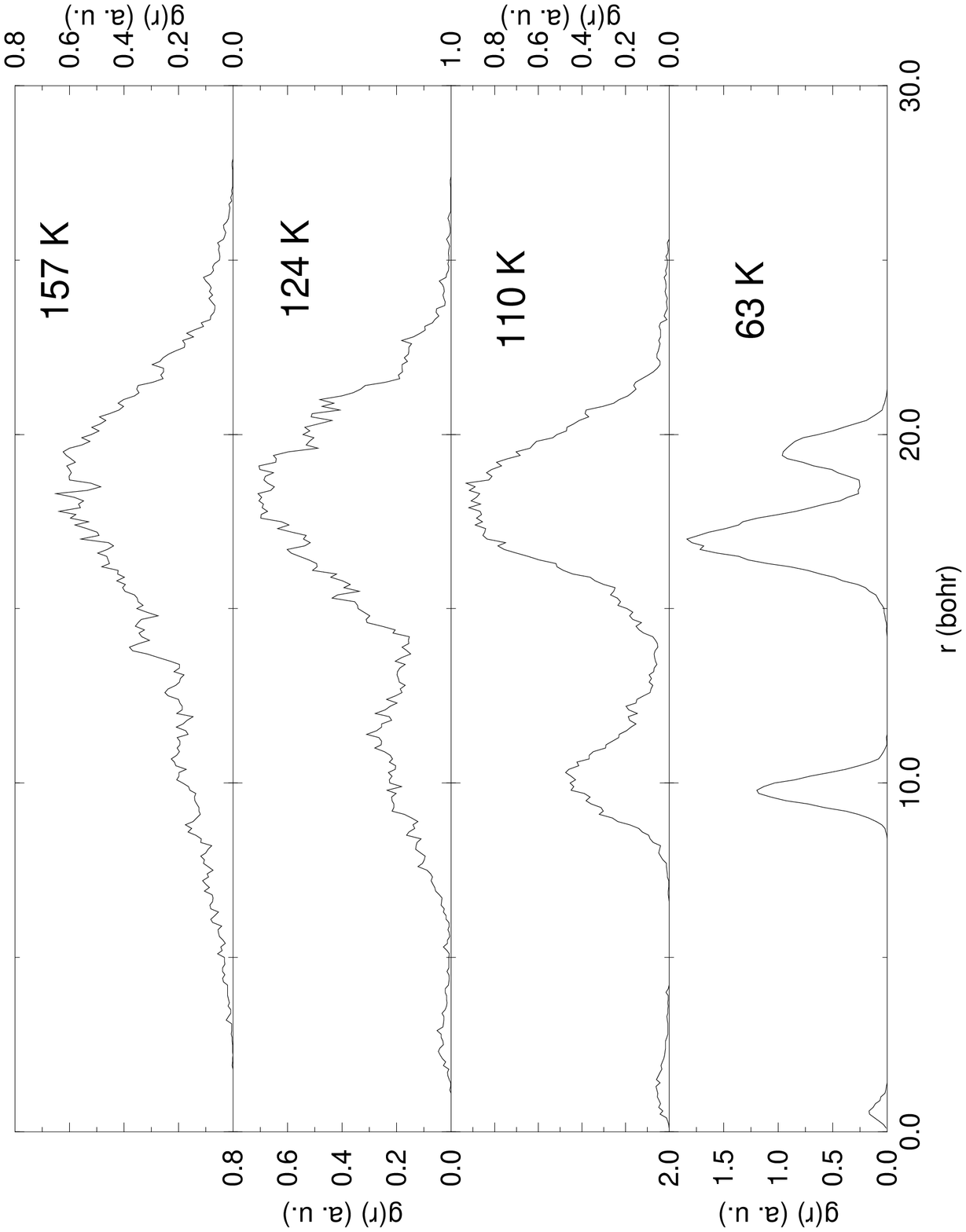}
\end{figure}
\end{document}